%% file: Granular-gripper-suction.tex
\documentclass[12pt,a4paper,twoside,twocolumn]{article}
\makeatletter\if@twocolumn\PassOptionsToPackage{switch}{lineno}\else\fi\makeatother

\usepackage{amsfonts,amssymb,amsbsy,latexsym,amsmath,tabulary,graphicx,times,xcolor}
\graphicspath{{figs/}}
\usepackage[T1]{fontenc}
\usepackage{url,multirow,morefloats,floatflt,cancel,tfrupee,multicol}
\makeatletter
\AtBeginDocument{\@ifpackageloaded{textcomp}{}{\usepackage{textcomp}}}
\makeatother
\usepackage{colortbl}
\usepackage{xcolor}
\usepackage{siunitx}
\usepackage{pifont}
\usepackage[nointegrals]{wasysym}
\usepackage[superscript,biblabel]{cite}
\makeatletter

\usepackage{pgfplots}
\usepackage{subcaption}
\usepackage{tikz}
\usetikzlibrary{arrows}
\pgfplotsset{compat=newest}
\usepgfplotslibrary{fillbetween}
\usetikzlibrary{math}
\usetikzlibrary{calc}
\usetikzlibrary{patterns}
\def\centerarc[#1](#2)(#3:#4:#5)
{ \draw[#1] ($(#2)+({#5*cos(#3)},{#5*sin(#3)})$) arc (#3:#4:#5); }
\usetikzlibrary{shapes.geometric}
\usetikzlibrary{positioning}
\usetikzlibrary{shapes}
\usetikzlibrary{intersections}
\usetikzlibrary{arrows.meta}
\tikzset{font=\small}
    \pgfdeclarepatternformonly{south west lines}{\pgfqpoint{-0pt}{-0pt}}{\pgfqpoint{3pt}{3pt}}{\pgfqpoint{3pt}{3pt}}{
        \pgfsetlinewidth{0.4pt}
        \pgfpathmoveto{\pgfqpoint{0pt}{0pt}}
        \pgfpathlineto{\pgfqpoint{3pt}{3pt}}
        \pgfpathmoveto{\pgfqpoint{2.8pt}{-.2pt}}
        \pgfpathlineto{\pgfqpoint{3.2pt}{.2pt}}
        \pgfpathmoveto{\pgfqpoint{-.2pt}{2.8pt}}
        \pgfpathlineto{\pgfqpoint{.2pt}{3.2pt}}
        \pgfusepath{stroke}}

\@ifundefined{etal}{\def\etal{\textit{et al.}}}{}

\AtBeginDocument{
\expandafter\ifx\csname eqalign\endcsname\relax
\def\eqalign#1{\null\vcenter{\def\\{\cr}\openup\jot\m@th
  \ialign{\strut$\displaystyle{##}$\hfil&$\displaystyle{{}##}$\hfil
      \crcr#1\crcr}}\,}
\fi
}

\@ifundefined{tsGraphicsScaleX}{\gdef\tsGraphicsScaleX{1}}{}
\@ifundefined{tsGraphicsScaleY}{\gdef\tsGraphicsScaleY{.9}}{}
\def\checkGraphicsWidth{\ifdim\Gin@nat@width>\linewidth
	\tsGraphicsScaleX\linewidth\else\Gin@nat@width\fi}

\def\checkGraphicsHeight{\ifdim\Gin@nat@height>.9\textheight
	\tsGraphicsScaleY\textheight\else\Gin@nat@height\fi}

\def\fixFloatSize#1{}
\let\ts@includegraphics\includegraphics

\def\inlinegraphic[#1]#2{{\edef\@tempa{#1}\edef\baseline@shift{\ifx\@tempa\@empty0\else#1\fi}\edef\tempZ{\the\numexpr(\numexpr(\baseline@shift*\f@size/100))}\protect\raisebox{\tempZ pt}{\ts@includegraphics{#2}}}}

\AtBeginDocument{\def\includegraphics{\@ifnextchar[{\ts@includegraphics}{\ts@includegraphics[width=\checkGraphicsWidth,height=\checkGraphicsHeight,keepaspectratio]}}}

\DeclareMathAlphabet{\mathpzc}{OT1}{pzc}{m}{it}

\def\URL#1#2{\@ifundefined{href}{#2}{\href{#1}{#2}}}

\edef\fntEncoding{\f@encoding}

\makeatother

\newif\ifmultipleabstract\multipleabstractfalse%
\usepackage{enumitem}

\usepackage[hmargin=1.8cm,vmargin=2.2cm]{geometry}
\usepackage{caption}
\usepackage{subcaption}

\makeatletter
\def\author#1{\gdef\@author{\hskip-\dimexpr(\tabcolsep)\hskip1pt\parbox{\dimexpr\textwidth-1pt}{\centering #1}}}

\let\@articletype\@empty \def\articletype#1{\gdef\@articletype{{\fontsize{14}{16}\selectfont #1}}}

\usepackage{fancyhdr}

\fancypagestyle{headings}{\fancyhf{}\fancyhead[RE]{\MakeTextUppercase{\textbf{\RunningAuthor}}}\fancyhead[LE]{\textbf{\thepage}}\fancyhead[LO]{\MakeTextUppercase{\textbf{\RunningHead}}}\fancyhead[RO]{\textbf{\thepage}}}
\fancypagestyle{plain}{\fancyhf{}\fancyfoot[C]{\textbf{\thepage}}}

\AtBeginDocument{\usepackage{lastpage}}
\def\title#1{%
  \gdef\@title{%
    \ifx\@articletype\@empty\else\@articletype~\\\fi%
     #1}%
}

\usepackage{abstract}
\def\abstractname{\textbf{Abstract}}

\def\NormalBaseline{\def\baselinestretch{1.1}}

\usepackage{textcase}
\usepackage[explicit]{titlesec}
\titleformat{\section}[block]{\NormalBaseline\boldmath\bfseries}
{\thesection.}
{6pt}
{#1}
[]
\titleformat{\subsection}[hang]{\NormalBaseline\filright\itshape}
{\thesubsection.}
{6pt}
{#1}
[]
\titleformat{\subsubsefction}[runin]{\NormalBaseline\filright\itshape}
{\hspace{16pt}\thesubsubsection}
{6pt}
{#1}
[]
\titleformat{\paragraph}[runin]{\NormalBaseline}
{\theparagraph}
{6pt}
{#1}
[]
\titleformat{\subparagraph}[runin]{\NormalBaseline}
{\thesubparagraph}
{6pt}
{#1}
[]

\titlespacing{\section}{0pt}{1.5\baselineskip}{.2\baselineskip}  
\titlespacing{\subsection}{0pt}{1.5\baselineskip}{.2\baselineskip}  
\titlespacing{\subsubsection}{0pt}{1.5\baselineskip}{.2\baselineskip}  
\titlespacing{\paragraph}{0pt}{.5\baselineskip}{10pt}  
\titlespacing{\subparagraph}{0pt}{.5\baselineskip}{10pt}

\usepackage{caption}
\usepackage{orcidlink}
\usepackage{subcaption}
\usepackage{lipsum}
\DeclareCaptionLabelFormat{figlabel}{Figure #2}
\date{}
\makeatother

\usepackage{float}

\graphicspath{{figs/}}

\begin{document}
\definecolor{persiangreen}{rgb}{0.0, 0.65, 0.58}

\title{Granular jamming gripper with integrated suction}
\def\RunningHead{
granular jamming gripper with integrated suction
}
\def\RunningAuthor{santarossa \etal}
\author{Angel Santarossa\orcidlink{0000-0002-7898-2986},
Olfa D'Angelo\orcidlink{0000-0002-7218-4596},
Achim Sack\orcidlink{0000-0003-2536-2879},
and Thorsten P\"oschel\orcidlink{0000-0001-5913-1070}
\thanks{Angel Santarossa, Olfa D'Angelo, Achim Sack, and Thorsten P\"oschel are with Institute for Multiscale Simulations,
        Friedrich-Alexander-Universit\"at Erlangen-N\"urnberg, Cauerstra\ss{}e 3, 91058 Erlangen, Germany
        {\tt\small thorsten.poeschel@fau.de}}%
}

\twocolumn[
\begin{@twocolumnfalse}
\maketitle
{\begin{abstract}
Granular grippers can manipulate a wide variety of objects, but need to be pressed on the object to conform to it. If the object is placed on unstable ground, e.g., on sand or water, this step might cause the object to sink or move away from the gripper, hindering proper operation. We introduce a granular gripper with an integrated suction cup, where suction and jamming are controlled independently. We demonstrate the system's robust and enhanced gripping capabilities by comparing its grasping performance with a typical granular gripper design. We show that the proposed device can grip objects that are challenging for typical granular grippers, including those placed on unstable ground, as the suction cup stabilizes the object, allowing the gripper to conform. 



\def\keywordstitle{Keywords}
\end{abstract}}
\end{@twocolumnfalse}
]

{
  \renewcommand{\thefootnote}%
    {\fnsymbol{footnote}}
  \footnotetext[1]{Angel Santarossa, Olfa D'Angelo, Achim Sack, and Thorsten P\"oschel are with Institute for Multiscale Simulations,
        Friedrich-Alexander-Universit\"at Erlangen-N\"urnberg, Cauerstra\ss{}e 3, 91058 Erlangen, Germany
        {\tt\small thorsten.poeschel@fau.de}
}
 


\section{Introduction}
Granular grippers are soft robotic effectors that exploit the effect of granular jamming to grip objects \cite{brown2010universal}. This type of gripper is highly adaptable and can easily manipulate objects of diverse geometry and surface properties, without the need for reconfiguration between gripping cycles\cite{amend2012positive,kapadia2012design,amend2016soft,miettinen2019granular,fitzgerald2020review}. The typical design of a granular gripper comprises a loosely packed granulate contained in a flexible, airtight membrane.
In this unjammed state, the granular material easily flows when deformed, akin to a fluid. When pressed against an object, the gripper deforms and conforms to the object's shape. To grip the object, the air is evacuated from within the gripper, which causes the gripper membrane to contract and compress the particles: the granular material jams and thus becomes mechanically stable by entering its solid-like state, characterized by a finite elastic modulus. 
In this jammed state, the gripper exerts forces on the target object that can effectively grip and firmly hold it. To release the object, the pressure inside the gripper 
is equalized to the ambient pressure, causing the membrane to relax and the granulate to return to a fluid-like state. 

The holding force generated by a granular gripper results from the combination of three mechanisms \cite{brown2010universal}: \emph{friction} due to tangential stress at the contact between the object and the membrane \cite{gomez2021effect,gotz2022soft}, \emph{interlocking} due to geometric constraints between the object and the jammed granulate \cite{kapadia2012design}, and \emph{suction} in regions where the membrane seals the surface of the object airtight \cite{santarossa2023effect}.

A key feature of granular jamming-based grippers is that they must be pressed against an object to conform to its shape \cite{amend2012positive,kremer2023trigger,gomez2021effect}. The maximum pushing force the gripper applies to the object is the \emph{activation force}. Deforming the gripper by compressing it onto an object works well for objects located on solid, stable ground. It works less reliably for objects in unstable positions such on rough surfaces, sand, wet soil, or floating on water \cite{licht2018partially}, where the object can sink or slide away from the gripper, see Fig. \ref{fig:operation}. In such situations, granular grippers tend to fail taking hold of objects or conform insufficiently to the object, resulting in low holding force.

\begin{figure}[htb]
    \centering
    \input{figs/typical-gripper-operation.tex}
    \caption{Granular gripper reaching out to get hold of an object floating on a liquid. The gripper fails because it cannot exert sufficient force on the evading object.}\label{fig:operation}
\end{figure}
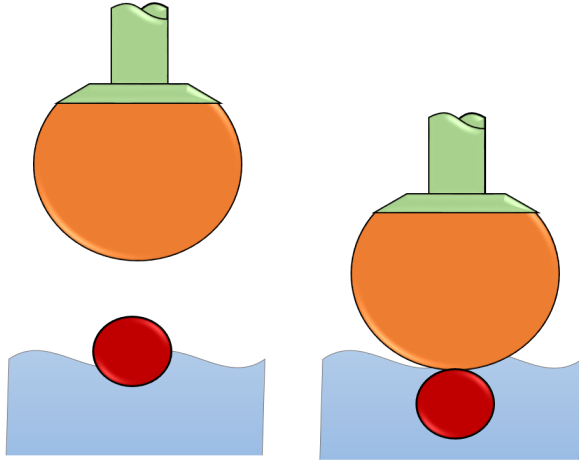

In addition, typical granular grippers cannot reliably manipulate objects whose shape or size impedes the gripper to cover a sufficiently large area. This is frequently the case for large objects, e.g., glass panes. Also problematic are very soft objects (e.g., dough, flexible foil, a plastic bag) that deform when in contact with the gripper \cite{brown2010universal}. 

We introduce a granular jamming gripper with integrated suction where suction- and jamming-based gripping can be activated independently. Granular gripper membranes equipped with suction cups have been proposed for grippers previously. In contrast to our device, however, either only suction was utilized for gripping \cite{tomokazu2015vacuum, takahashi2013development, gilday2023xeno, gilday2020suction}, or suction and granular jamming were activated simultaneously during gripping \cite{takahashi2016octopus}.
We demonstrate the performance of our gripper when applied to objects of different geometries (notably a large flat plate), sizes, and surface properties. We also demonstrate its ability to grip objects lying on unstable grounds and floating on water.

\section{Granular gripper with suction: design}

Our device is an ordinary granular gripper supplemented with a suction cup connected to an independent vacuum pump, as shown in Fig. \ref{fig:new-gripper}.
\begin{figure}[htb]
    \centering
    \input{figs/new-gripper-tikz2}
    \caption{Cross section of the gripper with suction mechanism.}
    \label{fig:new-gripper}
\end{figure}
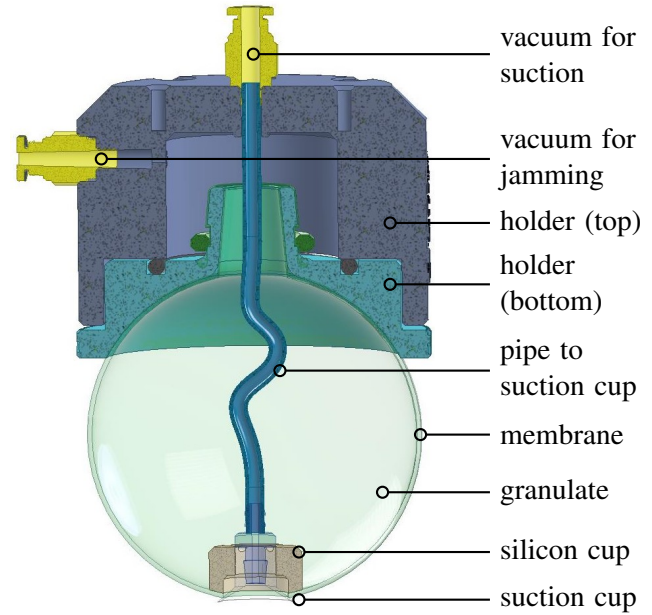

As for the typical gripper design, a holder maintains an elastic bag filled with particles. The gripper holder comprises two parts (blue and cyan in Fig. \ref{fig:new-gripper}). The membrane is attached to the bottom part of the holder and fixed with a snap ring. A plastic filter on top of the membrane retains the particles during the evacuation. An O-ring between the bottom and top part of the holder ensures proper sealing. 
At the side of the top part of the gripper holder, a flange allows loading the gripper's interior with positive air pressure or applying vacuum.

A suction cup is integrated into the lower center of the gripper membrane (Fig. \ref{fig:new-gripper}a). The suction cup is placed in a silicon cup to prevent the particles from deforming the suction cup during conformation. A flexible pipe connects the suction cup to a vacuum pump.

Note that suction and jamming-based gripping can be activated individually, which is crucial to grip and hold objects on unsteady ground.

\section{Experimental setup and procedures}

\subsection{Gripping objects on stable grounds}
The experimental setup consists of a fixed granular gripper and an object placed below, on a stage moving it along the $z$-direction, as sketched in Fig. \ref{fig:setup}. 

\begin{figure}[htb]
\centering
    \input{figs/setup}
    \caption{\label{fig:setup}Experimental setup for gripping objects from stable ground. The $z$-stage moves vertically. The object is attached to the $z$-stage via a load cell to measure the holding force.}
\end{figure}
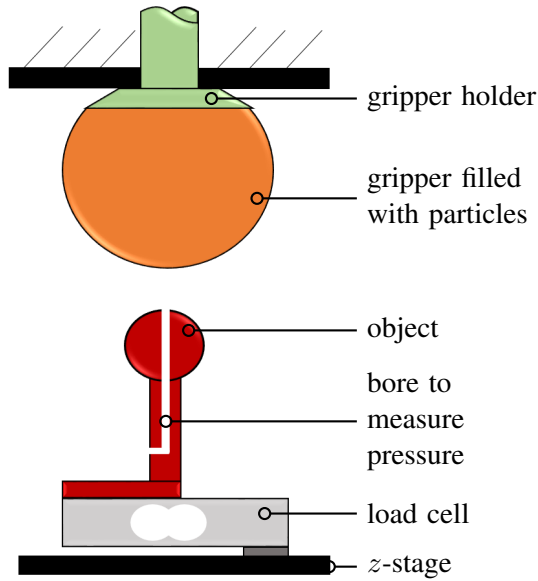

The elastic, non-permeable, spherical-shaped bag (diameter $73.0 \pm \SI{0.5}{\mm}$) of both grippers is filled with expanded polystyrene (EPS) beads (diameter $4.2 \pm \SI{0.5}{\mm}$). We use EPS beads because soft particles produce significantly higher holding forces in granular grippers \cite{gotz2022soft}. The particles are coated with an anti-static spray to minimize electrostatic interactions. The membrane is attached to the gripper holder.
The target object is mounted underneath on a $z$-stage. The $z$-stage position can be changed vertically using two stepper motors driven by a microcontroller. A load cell between the object and the $z$-stage measures the force acting on the target object throughout the gripping process. 

In the following, we compare the granular gripper with suction to an identical device not equipped with a suction cup, referred to here as ``ordinary gripper''. We study the gripping of six 3D-printed objects (three shapes, each with a smooth and rough surface): 
\begin{itemize}
\item a small sphere (diameter $20.0 \pm \SI{0.5}{\mm}$, mass $17.5 \pm \SI{0.1}{\gram}$)
\item a large sphere (diameter $40.0 \pm \SI{0.5}{\mm}$, mass $23.2 \pm \SI{0.1}{\gram}$ )
\item a large square flat plate (surface $2500 \pm \SI{1}{\square\mm}$, thickness $6.0 \pm \SI{0.5}{\mm}$,  mass $26.2 \pm \SI{0.1}{\gram}$) 
\end{itemize}
To obtain smooth surfaces, the objects are coated with lacquer (Lackspray matt Acryl) to facilitate suction between the membrane and the object. For rough surfaces, a mesh structure was printed on the surfaces of the objects to prevent suction. All objects have a borehole at the top, connected to a pressure sensor to monitor the pressure at the membrane-object interface (see Fig. \ref{fig:setup}). We studied (i) the ordinary granular gripper, (ii) our gripper with only suction, and (iii) our gripper with a combination of suction and jamming. Suction is active throughout each measurement. A pressure difference  $p_\text{suc}\approx \SI{90}{\kilo\pascal}$ between ambient and the suction cup is applied for suction.

Before each measurement, the interior of the gripper is pressurized at $\SI{5}{\kilo\pascal}$ to fluidize the enclosed granulate and erase the memory of previous cycles. An experimental measurement comprises the following steps: 
\begin{enumerate}[label=(\roman*)]
\item The $z$-stage displaces the object upwards, pushing it onto the gripper. 
\item When a predefined indentation depth is reached, the platform stops. Then the granulate relaxes for \SI{20}{\s}. In the first two steps, the gripper adapts to the shape of the object.
\item The air is evacuated from the gripper. The pressure difference between the interior of the gripper and the surrounding atmosphere ($p_\text{vac}\approx \SI{90}{\kilo\pascal}$) causes the granulate to jam. 
\item The platform moves downwards until the object is no longer in contact with the gripper, which imitates the manipulation of an object. 
\end{enumerate}

\subsection{Gripping on unstable grounds}

To test the gripping performance of objects placed on unstable ground, the object previously fixed to a $z$-stage is replaced by an open-top container of length \SI{56.0}{\mm}, width \SI{70.0}{\mm} and height \SI{41.0}{\mm}, filled with water or sand, as shown in Fig. \ref{fig:setup2}. 
A spherical object is attached to the base of the box with a thread in such a way that the object can be moved by a distance that equals the height of the box. When the object is pulled to the maximum length of the thread, it is entirely outside of the water or sand. 

As the object is pulled out but remains connected to the box and the force sensor, the maximum holding force is measured similarly to the previous setup (Fig. \ref{fig:setup}).

The spherical object consists of two parts: a lower smaller one and an upper larger one. 
The upper part is hollow (lower density), and the lower one is solid (higher density), so the object floats on water. The box is filled up to 90\% of its volume with water or sand, and the test object is placed on the surface. 
\begin{figure}[htb]
\centering
    \input{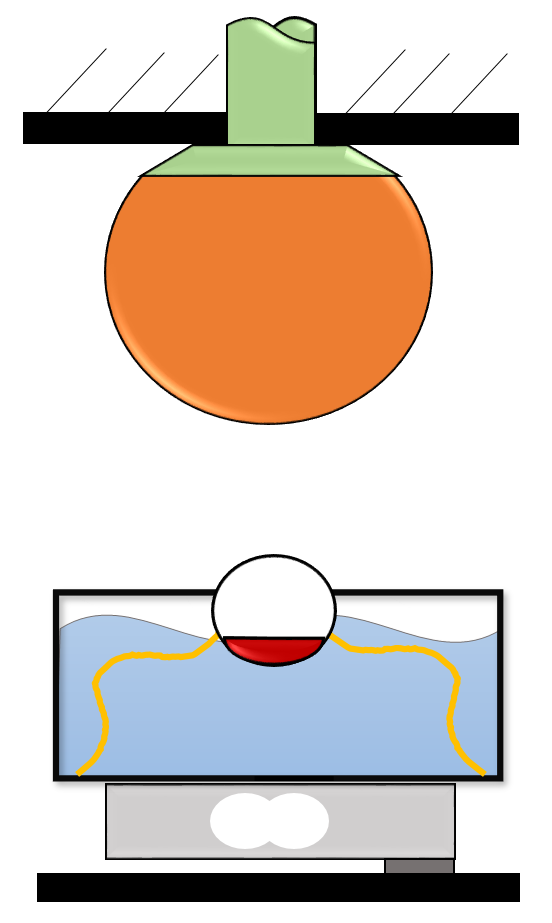}
    \caption{\label{fig:setup2} Gripping objects from unstable grounds. An open box, placed on the $z$-stage, is filled with water or sand. The test object is placed on the unstable ground's surface and connected to the bottom of the box by a thread. The test object comprises two parts: a lower solid one and an upper hollow one. The difference in densities allows the object to float in water. The box is attached to a load cell to record the holding force.}
\end{figure}

\section{Results}
\subsection{Gripping from stable grounds}
We first study the effect of adding suction to a granular gripper for manipulating objects of various shapes and surface properties placed on stable grounds. Figure \ref{fig:HF1} shows the maximum holding force and maximum differential pressure (difference between the ambient pressure and the pressure at the interface between the membrane and the object), $\Delta p$, for the studied cases.

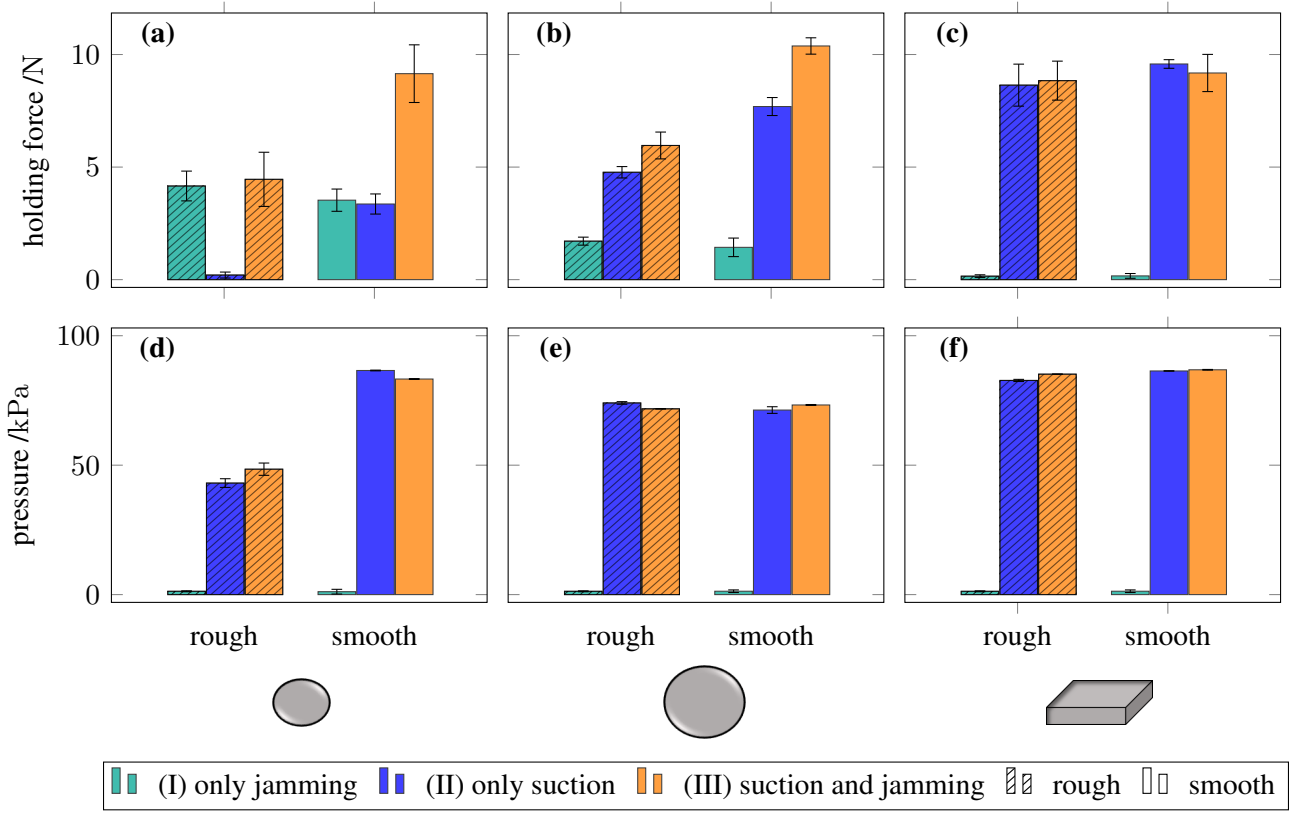
\begin{figure*}
\centering\hspace{-54mm}
\input{figs/hf-plate}
\vspace{8mm}
\caption{\label{fig:HF1}
Maximum holding force (a-c) and maximum differential pressure, $\Delta p$ (d-e) for the studied cases. Colors indicate the employed gripping mechanisms: (I) ordinary granular gripper (jamming only), (II) our gripper where only suction is active (no jamming), and (III) our gripper with both mechanism, suction and jamming, activated. All data are shown for smooth and rough surfaces.
Error bars indicate the standard deviation due to six independent measurements; the heights of the columns show the mean.
}
\end{figure*}

For a small sphere whose diameter is less than half the gripper's diameter (Fig. \ref{fig:HF1}a, d),  the holding force is ca. \SI{4}{\newton} for both rough and smooth surface, when exploiting jamming only (case (I)), which is sufficient for reliable operation. Using only the suction cup (case (II)), 
for the smooth surface, we reach up to $\approx$\SI{3.4}{\newton}, but on the rough surface, the gripper fails, evidenced by minimal holding force of $\approx$\SI{0.2}{\newton}. 
For suction and jamming both activated (case (III)), we obtain the largest forces, ca. \SI{4.5}{\newton} for the rough sphere and ca. \SI{9.2}{\newton} for the smooth one which is an improvement by ca. the factor $2.5$ compared to the ordinary gripper.

For the large sphere whose diameter is more than half the gripper's diameter (Fig.~\ref{fig:HF1}b, e), the holding force of an ordinary gripper (case (I)) is low for both smooth and rough surface (ca.~\SI{1.5}{\newton}) since the gripper cannot embrace the sphere. No sealed cavities appear at the object-gripper interface, apparent from minimal pressure difference (Fig. \ref{fig:HF1}e). When only suction is active (case (II)), the holding force is ca. \SI{4.8}{\newton} for a rough surface and ca. \SI{7.7}{\newton} for a smooth surface. Uneven surfaces lead to a smaller force, as roughness weakens the seal between the object and the suction cup. 
For both mechanisms active, jamming and suction (case (III)), the largest holding forces are achieved, ca. \SI{6.0}{\newton} for a rough surface and \SI{10.4}{\newton} for a smooth one.

For a flat pane larger than the gripper (Fig. \ref{fig:HF1}c, f), the ordinary gripper fails for both surfaces (force $<$\SI{0.2}{\newton}), which is expected: objects that the gripper cannot embrace due to their shape or size are challenging for granular grippers \cite{brown2010universal}. In contrast, significant holding forces (ca. \SI{9}{\newton}) are achieved by the gripper with integrated suction, irrespective of the object's surface properties (rough or smooth). For both cases (II) and (III), i.e., suction only or suction and jamming, we obtain similar values for the holding forces and the differential pressure, which indicates that jamming does not make a significant contribution since the gripper fails to embrace the object.

For all measurements obtained for the ordinary granular gripper, no appreciable pressure difference, $\Delta p$, is measured at the membrane-object interface (case (I) in Fig.~\ref{fig:HF1}). This indicates that airtight cavities are not formed between the object and the gripper; therefore, in contrast to earlier speculations \cite{santarossa2023effect}, suction due to jamming only (i.e., not due to an independent suction cup) is not achieved. 

For all objects tested, regardless of their surface state, jamming-based gripping in combination with an independent suction cup considerably increases the maximum holding force. For objects that are challenging for ordinary granular grippers (large sphere and flat plate), the ordinary gripper fails, while integrated suction effectively allows for proper operation.

\subsection{Gripping from unstable grounds}
To study the gripping of objects placed on unstable grounds, we perform the procedure sketched in Fig. \ref{fig:operation2}: 
\begin{figure*}[htb]
 \centering
     \input{figs/sketch-new-gripper-method}
    \caption{\label{fig:operation2}Sequence of steps for gripping a sphere (in red) floating on water with a granular gripper with integrated suction.}
\end{figure*}
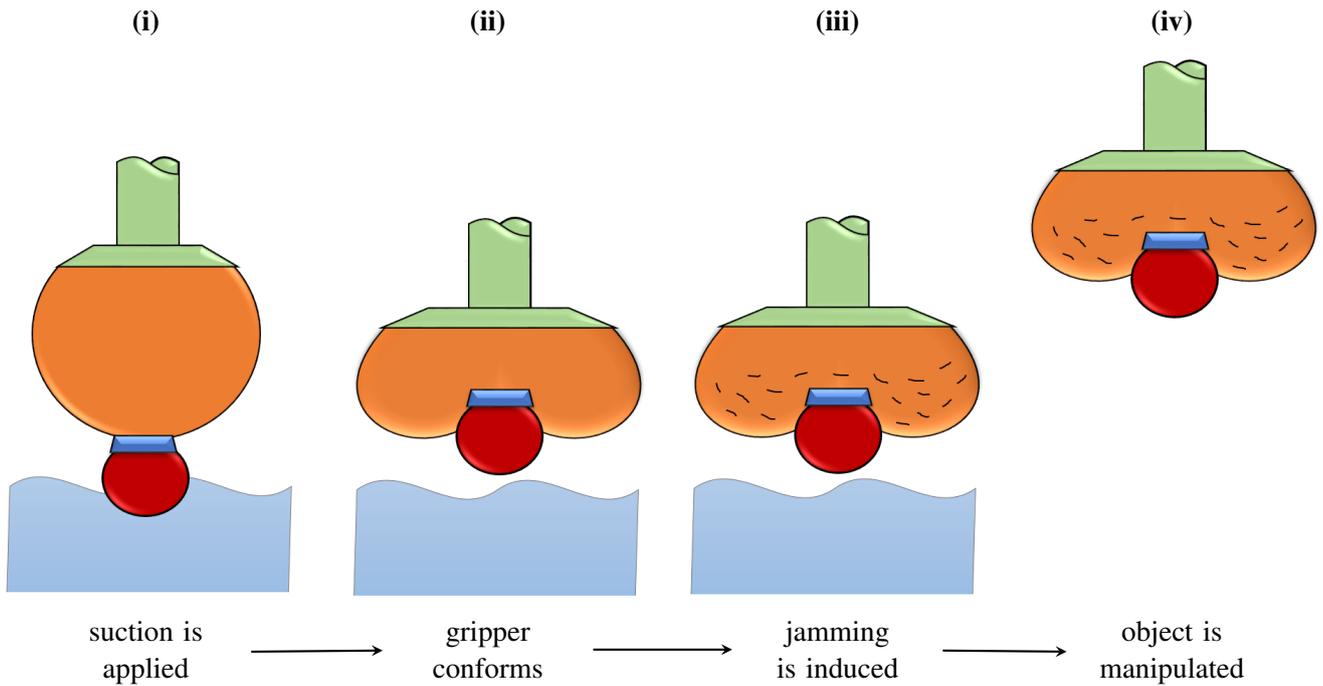
\begin{enumerate}[label=(\roman*)]
\item Suction is applied to fix the object's position, preventing it from sinking or sliding away from the gripper. 
\item Air pressure is applied within the gripper to fluidize the granulate and inflate the gripper's membrane, which increases the contact surface between the gripper and the object.
\item Applying vacuum enforces the grains to jam, causing the gripper to clamp the object. 
\item The object can then be handled safely.
\end{enumerate}

Figure \ref{fig:HF2} shows the maximum holding force when a sphere of diameter $20.0 \pm \SI{0.5}{\mm}$ is gripped from a sand surface or floating on water. Again, we study the ordinary granular gripper (case (I)) versus a granular gripper with integrated suction (case (III)). 
\begin{figure}[b!]
     \begin{subfigure}{\columnwidth}
         \centering
         \input{figs/hf-water}
     \end{subfigure}
     \par\bigskip
     \begin{subfigure}{\columnwidth}
         \centering
         \input{figs/hf-sand}
     \end{subfigure}
        \caption{Maximum holding force for gripping a sphere from an unstable ground. (top) sphere is placed on a sand surface, (bottom) sphere floats on water. Data is shown for an ordinary granular gripper (case I) and a granular gripper with integrated suction (case III). Error bars show the standard deviation over six independent measurements; the height of the bar shows the mean. \label{fig:HF2}}
\end{figure}
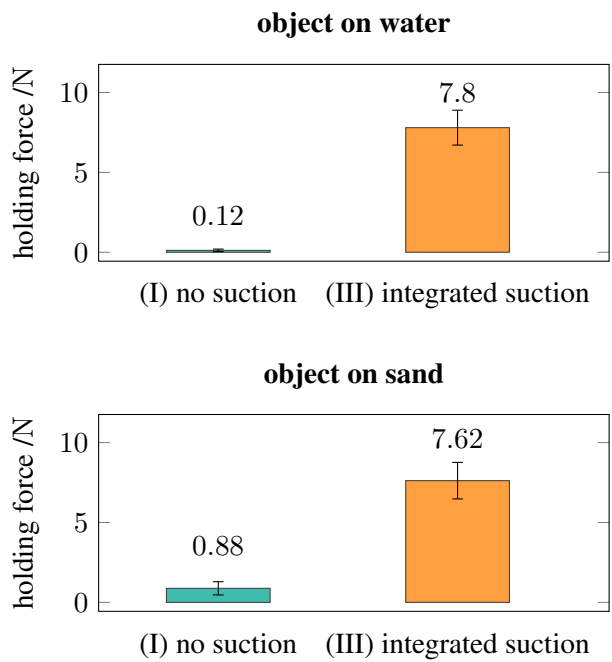

When an ordinary gripper is pressed against the floating sphere to conform to its shape, the object sinks. Therefore, the gripper embraces only a small part of the object, resulting in a low holding force of ca. \SI{0.1}{\newton}. Effectively, the gripper cannot grip the object, as it slips out of grip almost immediately. 

For activated suction, the object is held in place during conformation. When jamming is enforced, the holding force reaches \SI{7.8}{\newton}, corresponding to an increase of about one order of magnitude compared to the ordinary gripper. 

Similarly, when the sphere is placed on a sand surface, it tends to sink into the ground when an ordinary granular gripper is pressed against its surface. Consequently, the conformation is hindered, and only a reduced holding force (ca. \SI{1}{\newton}) is achieved, see Fig. \ref{fig:HF2} (bottom). Using the gripper with activated suction (case (III)), a significantly larger holding force of \SI{7.62}{\newton} was found.

The results reveal that applying an independent suction mechanism can considerably enhance the performance of granular grippers. In particular, such grippers can grip objects located on unstable ground or even floating on water, which is impossible for ordinary granular grippers.

\section{Summary}
Granular grippers are capable of gripping objects of different shapes and sizes. Despite their versatility, they cannot reliably grip flat, very soft, or deformable objects, objects larger than the gripper size, and objects resting on unstable ground. This paper presents a granular gripper with an integrated suction mechanism. 

We show that this gripper performs better gripping and holding than ordinary granular grippers. For objects larger than the gripper and flat objects, adding an independent suction mechanism increases the holding force by about an order of magnitude. The improved gripper can also grip objects located on unstable grounds, such as water or sand, where ordinary granular grippers fail.

Granular grippers with additional suction devices can be used in demanding applications, e.g., picking up and placing objects of different geometries in factory automation or improving prosthetic systems. Due to its robustness and ability to grip objects on unstable surfaces, the presented gripper can be integrated into remote-controlled flying devices \cite{kremer2023trigger}. 

\section*{Acknowledgements} 
 The authors thank Walter Pucheanu for technical support. This work was supported by the Interdisciplinary Center for Nanostructured Films (IZNF), the Competence Unit for Scientific Computing (CSC), and the Interdisciplinary Center for Functional Particle Systems (FPS) at Friedrich-Alexander-Universit\"at Erlangen-N\"urnberg.

\section*{Funding Statement}
Angel Santarossa and Thorsten P\"oschel gratefully acknowledge funding by Deutsche Forschungsgemeinschaft (DFG, German Research Foundation)--Project Number
411517575.

\section*{Author Disclosure Statement}
The authors declare that they have no conflict of interest.

\bibliographystyle{vancouver}
\bibliography{Gripper-suction.bib}

\end{document}

%% file: figs/typical-gripper-operation.tex
\begin{tikzpicture}
\newcommand{\FixedLengthArrow}{2.5,0}

\node[] (setup)  at (0, 0) {}; 








\node[below = 0mm of setup, right= 0mm, anchor=north west] (setupwater) {\includegraphics[width=0.9\linewidth]
{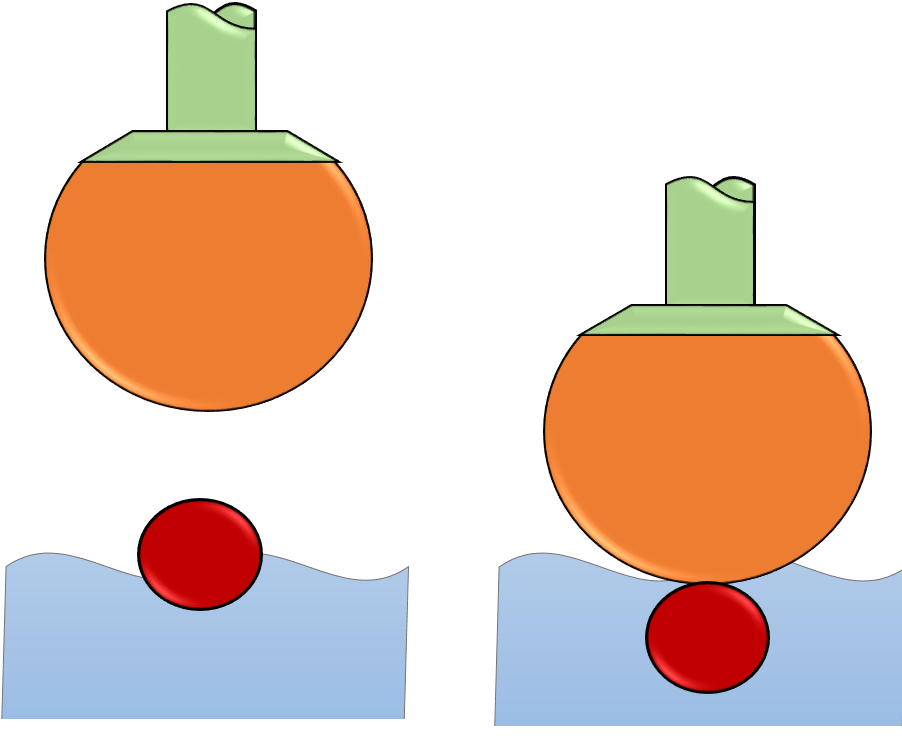}};





\end{tikzpicture}

%% file: figs/new-gripper-tikz2.tex
\begin{tikzpicture}
\small

\def\widthfortext{2}
\node (view1) at (0, 0) {\includegraphics[width=0.32\textwidth]
{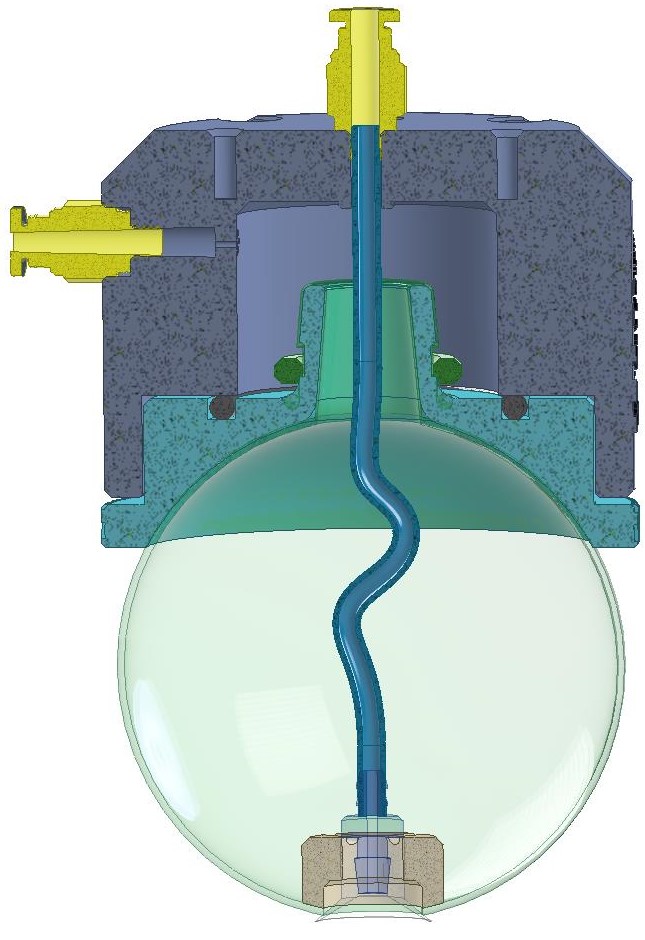}};


\node[right=6mm of view1.6, anchor=west, text width=\widthfortext cm,] (vacuum) 
{holder \\ (bottom)};
\node[left=14mm of vacuum] (vacuumPt) {};
\draw[-{Circle[fill=none]}, thick](vacuum.west)--(vacuumPt);

\node[right=6mm  of view1.21, anchor=west,text width=\widthfortext cm] (vacjam) 
{holder (top) };
\node[left=14mm of vacjam] (vacjamPt) {};
\draw[-{Circle[fill=none]}, thick](vacjam.west)--(vacjamPt);

\node[right=6mm  of view1.34, anchor=west,text width=\widthfortext cm] (vacjam) 
{vacuum for \\jamming };
\node[left=52mm of vacjam] (vacjamPt) {};
\draw[-{Circle[fill=none]}, thick](vacjam.west)--(vacjamPt);

\node[right=6mm  of view1.49, anchor=west,text width=\widthfortext cm] (vacjam) 
{vacuum for \\ suction };
\node[left=32mm of vacjam] (vacjamPt) {};
\draw[-{Circle[fill=none]}, thick](vacjam.west)--(vacjamPt);


\node[right=6mm of view1.-16, anchor=west, text width=\widthfortext cm] (cupcon) 
{pipe to \\ suction cup};
\node[left=28.5mm of cupcon] (cupconPt) {};
\draw[-{Circle[fill=none]}, thick](cupcon.west)--(cupconPt);

\node[right=6mm of view1.-30, anchor=west, text width=\widthfortext cm] (membrane) 
{membrane};
\node[left=10mm of membrane] (membranePt) {};
\draw[-{Circle[fill=none]}, thick](membrane.west)--(membranePt);

\node[right=6mm of view1.-40, anchor=west, text width=\widthfortext cm] (granu) 
{granulate};
\node[left=15mm of granu] (granuPt) {};
\draw[-{Circle[fill=none]}, thick](granu.west)--(granuPt);

\node[right=6mm of view1.-48, anchor=west, text width=\widthfortext cm] (silcup) 
{silicon cup};
\node[left=26mm of silcup] (silcupPt) {};
\draw[-{Circle[fill=none]}, thick](silcup.west)--(silcupPt);

\node[right=6mm of view1.-53, anchor=west,text width=\widthfortext cm] (cup)
{suction cup};
\node[left=26mm of cup] (cupPt) {};
\draw[-{Circle[fill=none]}, thick](cup.west)--(cupPt);


\end{tikzpicture}

%% file: figs/setup.tex
\begin{tikzpicture}
\small 
\node[] (setup)  at (0, 0) { \includegraphics[width=.6\columnwidth]{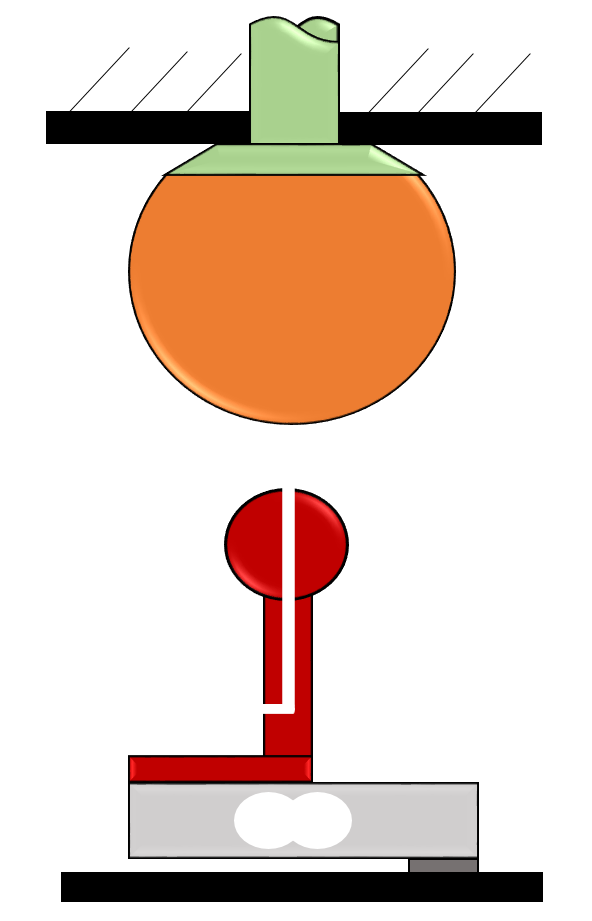}};

\node[right=-2mm of setup.44, anchor=west, text width=2.5cm, text=black] (holder) {gripper holder};
\node[left=20mm of holder] (holderPt) {};
\draw[-{Circle[fill=none]}, thick](holder.west)--(holderPt);

\node[right=-2mm of setup.25, anchor=west, text width=2.5cm, text=black] (gripper) 
{gripper filled \\with particles};
\node[left=14mm of gripper] (gripperPt) {};
\draw[-{Circle[fill=none]}, thick](gripper)--(gripperPt);

\node[right=-2mm of setup.-11, anchor=west, text width=2.5cm, text=black] (object) {object};
\node[left=23mm of object] (objectPt) {};
\draw[-{Circle[fill=none]}, thick](object.west)--(objectPt);

\node[right=-2mm of setup.-33, anchor=west, text width=2.5cm, text=black] (bore) {bore to \\ measure \\ pressure};
\node[left=26mm of bore] (borePt) {};
\draw[-{Circle[fill=none]}, thick](bore.west)--(borePt);

\node[right=-2mm of setup.-48, anchor=west, text width=2.5cm, text=black] (loadcell) {load cell};
\node[left=13mm of loadcell] (loadcellPt) {};
\draw[-{Circle[fill=none]}, thick](loadcell.west)--(loadcellPt);


\node[right=-2mm of setup.-54, anchor=west,text width=2.5cm, text=black] (zstage) {$z$-stage};
\node[left=4.5mm of zstage] (zstagePtmid) {};
\draw[-{Circle[fill=none]}, thick](zstage.west)--(zstagePtmid); 

\end{tikzpicture}

%% file: figs/setup2-corrected.tex
\begin{tikzpicture}
\small 
\node[] (setup)  at (0, 0) { \includegraphics[width=.67\columnwidth]{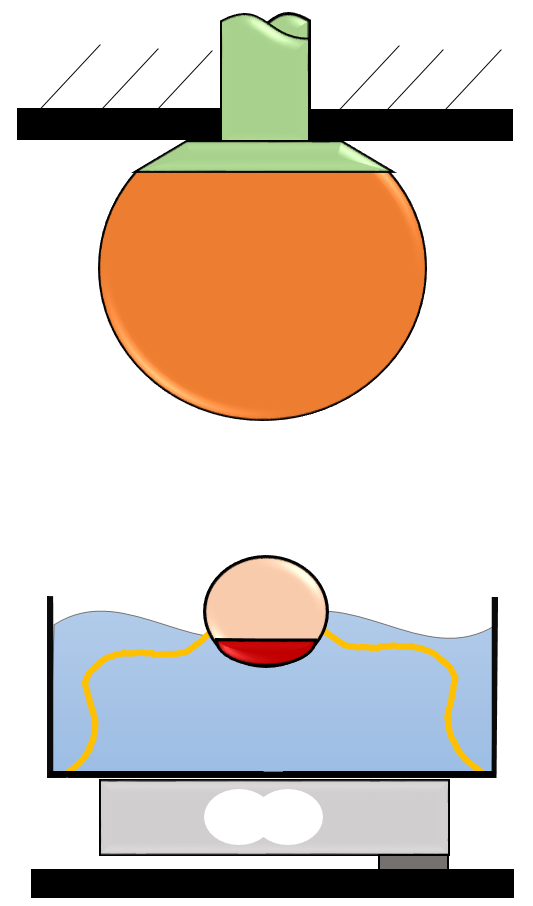}};

\node[right=1mm of setup.47, anchor=west, text width=2.5cm, text=black] (holder) {gripper holder};
\node[left=24mm of holder] (holderPt) {};
\draw[-{Circle[fill=none]}, thick](holder.west)--(holderPt);

\node[right=1mm of setup.25, anchor=west, text width=2.5cm, text=black] (gripper) 
{gripper filled \\with particles};
\node[left=17mm of gripper] (gripperPt) {};
\draw[-{Circle[fill=none]}, thick](gripper)--(gripperPt);

\node[right=1mm of setup.-24, anchor=west, text width=2.5cm, text=black] (object) 
{hollow part};
\node[left=31mm of object] (objectPt) {};
\draw[-{Circle[fill=none]}, thick](object.west)--(objectPt);

\node[right=1mm of setup.-35, anchor=west, text width=2.5cm, text=black] (object) 
{solid part};
\node[left=31mm of object] (objectPt) {};
\draw[-{Circle[fill=none]}, thick](object.west)--(objectPt);

\node[right=1mm of setup.-43, anchor=west, text width=2.5cm, text=black] (bore) {thread};
\node[left=12mm of bore] (borePt) {};
\draw[-{Circle[fill=none]}, thick](bore.west)--(borePt);

\node[right=1mm of setup.-47, anchor=west, text width=2.5cm, text=black] (water) {water or sand};
\node[left=22.5mm of water] (waterPt) {};
\draw[-{Circle[fill=none]}, thick](water.west)--(waterPt);

\node[right=1mm of setup.-53, anchor=west, text width=2.5cm, text=black] (loadcell) {load cell};
\node[left=14mm of loadcell] (loadcellPt) {};
\draw[-{Circle[fill=none]}, thick](loadcell.west)--(loadcellPt);


\node[right=1mm of setup.-57, anchor=west,text width=2.5cm, text=black] (zstage) {$z$-stage};
\node[left=5.0mm of zstage] (zstagePtmid) {};
\draw[-{Circle[fill=none]}, thick](zstage.west)--(zstagePtmid); 

\end{tikzpicture}

%% file: figs/hf-plate.tex
\begin{tikzpicture}

\pgfplotsset{set layers}
\definecolor{mediumpersianblue}{rgb}{0.0, 0.4, 0.65}
\definecolor{pastelorange}{rgb}{1.0, 0.7, 0.28}
\definecolor{orangepeel}{rgb}{1.0, 0.62, 0.0}
\definecolor{persiangreen}{rgb}{0.0, 0.65, 0.58}

\def\widthgraphs{0.375}
\def\heightgraphs{0.3}
\def\graphsxshift{0.3}
\def\graphsyshift{0.24}

    \begin{axis}[
     bar width=7pt,
    ybar,
    width=\widthgraphs\textwidth, 
    height=\heightgraphs\textwidth,
    ymin=0,ymax=11.5,
    enlarge y limits = 0.03,
    ylabel={holding force /\si{\newton}},
	xmin=0.0,xmax=5.0,
    legend image code/.code={%
      \draw[#1] (0cm,-0.1cm) rectangle (0.6cm,0.1cm);
    } 
	x tick style={draw=none},
    xtick={1.5,3.5},
	xticklabels={ },
	every node near coord/.style={
        opacity=1, 
        text depth=6.5mm,
        /pgf/number format/precision=2
        },
	every axis plot/.append style={
          ybar,
          bar width=.5,
          bar shift=0pt,
          fill
        },
    legend to name={mylegendA},
    ]

\addplot [fill = persiangreen, opacity=0.75,
    postaction={pattern=south west lines},
        error bars/.cd,
        y dir=both,
        y explicit relative, 
        error bar style={color=black, solid, opacity=1}
        ] 
coordinates {
    (1, 4.16) +- (0.159,0.159)
    };
\addplot [fill = persiangreen, opacity=0.75,
        error bars/.cd,
        y dir=both,
        y explicit relative, 
        error bar style={color=black, solid, opacity=1}
        ] 
coordinates {
    (3, 3.53) +- (0.14,0.14)
    };

\addplot [fill = blue, opacity=0.75,
    postaction={pattern=south west lines},
        error bars/.cd,
        y dir=both,
        y explicit relative, 
        error bar style={color=black, solid, opacity=1}
        ] 
coordinates {
    (1.515, 0.204) +- (0.65,0.65)
    };
\addplot [fill = blue, font=\scriptsize, opacity=0.75,
        error bars/.cd,
        y dir=both,
        y explicit relative, 
        error bar style={color=black, solid, opacity=1}
        ] 
coordinates {
    (3.515, 3.36) +- (0.133,0.133)
    };

\addplot [fill = orange, font=\scriptsize, opacity=0.75,
    postaction={pattern=south west lines},
        error bars/.cd,
        y dir=both,
        y explicit relative, 
        error bar style={color=black, solid, opacity=1}
        ] 
coordinates {
    (2.03, 4.456) +- (0.27,0.27)
	};
\addplot [fill = orange, font=\scriptsize, opacity=0.75,
        error bars/.cd,
        y dir=both,
        y explicit relative, 
        error bar style={color=black, solid, opacity=1}
        ] 
coordinates {
    (4.03, 9.15) +- (0.14,0.14)
	};

\node at (rel axis cs: .2,.92) [left] {\textbf{(a)}};
        
\end{axis}

\node at (2.5,-5.5) {\includegraphics[width=0.8cm]
{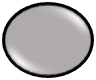}};


    \begin{axis}[
    xshift=\graphsxshift\textwidth,
     bar width=7pt,
    ybar,
    width=\widthgraphs\textwidth, 
    height=\heightgraphs\textwidth,
    ymin=0,ymax=11.5,
    enlarge y limits = 0.03,
    ylabel={},
    yticklabels={ },
	xmin=0.0,xmax=5.0,
    legend image code/.code={%
      \draw[#1] (0cm,-0.1cm) rectangle (0.6cm,0.1cm);
    } 
	x tick style={draw=none},
    xtick={1.5,3.5},
	xticklabels={ },
	every node near coord/.style={
        opacity=1, 
        text depth=6.5mm,
        /pgf/number format/precision=2
        },
	every axis plot/.append style={
          ybar,
          bar width=.5,
          bar shift=0pt,
          fill
        },
    legend to name={mylegendB},
    ]

\addplot [fill = persiangreen, opacity=0.75,
    postaction={pattern=south west lines},
        error bars/.cd,
        y dir=both,
        y explicit relative, 
        error bar style={color=black, solid, opacity=1}
        ] 
coordinates {
    (1, 1.71) +- (0.104,0.104)
    };
\addplot [fill = persiangreen, opacity=0.75,
        error bars/.cd,
        y dir=both,
        y explicit relative, 
        error bar style={color=black, solid, opacity=1}
        ] 
coordinates {
    (3, 1.435) +- (0.287,0.287)
    };
    
\addplot [fill = blue, font=\scriptsize, opacity=0.75,
    postaction={pattern=south west lines},
        error bars/.cd,
        y dir=both,
        y explicit relative, 
        error bar style={color=black, solid, opacity=1}
        ] 
coordinates {
    (1.515, 4.77) +- (0.053,0.053)
    };
\addplot [fill = blue, font=\scriptsize, opacity=0.75,
        error bars/.cd,
        y dir=both,
        y explicit relative, 
        error bar style={color=black, solid, opacity=1}
        ] 
coordinates {
    (3.515, 7.69) +- (0.052,0.052)
    };

\addplot [fill = orange, font=\scriptsize, opacity=0.75,
    postaction={pattern=south west lines},
        error bars/.cd,
        y dir=both,
        y explicit relative, 
        error bar style={color=black, solid, opacity=1}
        ] 
coordinates {
    (2.03, 5.96) +- (0.1,0.1)
	};
\addplot [fill = orange, font=\scriptsize, opacity=0.75,
        error bars/.cd,
        y dir=both,
        y explicit relative, 
        error bar style={color=black, solid, opacity=1}
        ] 
coordinates {
    (4.03, 10.38) +- (0.035,0.035)
	};        

\node at (rel axis cs: .2,.92) [left] {\textbf{(b)}};
        
    \end{axis}

\node at (7.8,-5.5) {\includegraphics[width=1.1cm]{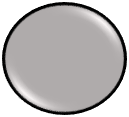}};


\begin{axis}[
    xshift=2*\graphsxshift\textwidth,
     bar width=7pt,
    ybar,
    width=\widthgraphs\textwidth, 
    height=\heightgraphs\textwidth,
    ymin=0,ymax=11.5,
    enlarge y limits = 0.03,
    ylabel={},
    yticklabels={ },
	xmin=0.0,xmax=5.0,
    legend entries={typical gripper, proposed gripper: suction, proposed gripper: jamming + suction},
    legend image code/.code={%
      \draw[#1] (0cm,-0.1cm) rectangle (0.6cm,0.1cm);
    } 
	x tick style={draw=none},
    xtick={1.5,3.5},
	xticklabels={ },
	every node near coord/.style={
        opacity=1, 
        text depth=6.5mm,
        /pgf/number format/precision=2
        },
	every axis plot/.append style={
          ybar,
          bar width=.5,
          bar shift=0pt,
          fill
        },
    legend to name={mylegendC},
    ]

\addplot [fill = persiangreen, font=\scriptsize, opacity=0.75,
    postaction={pattern=south west lines},
        error bars/.cd,
        y dir=both,
        y explicit relative, 
        error bar style={color=black, solid, opacity=1}
        ] 
coordinates {
    (1, 0.154) +- (0.40,0.40)
    };
\addplot [fill = persiangreen, font=\scriptsize, opacity=0.75,
        error bars/.cd,
        y dir=both,
        y explicit relative, 
        error bar style={color=black, solid, opacity=1}
        ] 
coordinates {
    (3, 0.166) +- (0.64,0.64)
    };

\addplot [fill = blue, font=\scriptsize, opacity=0.75,
    postaction={pattern=south west lines},
        error bars/.cd,
        y dir=both,
        y explicit relative, 
        error bar style={color=black, solid, opacity=1}
        ] 
coordinates {
    (1.515, 8.64) +- (0.108,0.108)
    };
\addplot [fill = blue, font=\scriptsize, opacity=0.75,
        error bars/.cd,
        y dir=both,
        y explicit relative, 
        error bar style={color=black, solid, opacity=1}
        ] 
coordinates {
    (3.515, 9.58) +- (0.02,0.02)
    };

\addplot [fill = orange, font=\scriptsize, opacity=0.75,
    postaction={pattern=south west lines},
        error bars/.cd,
        y dir=both,
        y explicit relative, 
        error bar style={color=black, solid, opacity=1}
        ] 
coordinates {
    (2.03, 8.84) +- (0.098,0.098)
	};
\addplot [fill = orange, font=\scriptsize, opacity=0.75,
        error bars/.cd,
        y dir=both,
        y explicit relative, 
        error bar style={color=black, solid, opacity=1}
        ] 
coordinates {
    (4.03, 9.18) +- (0.09,0.09)
	};

\node at (rel axis cs: .2,.92) [left] {\textbf{(c)}};
        
    \end{axis}

\node at (13.,-5.5) {\includegraphics[width=1.4cm]{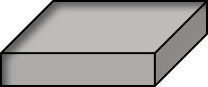}};


\begin{axis}[
    bar width=7pt,
    ybar,
    yshift=-\graphsyshift\textwidth,
    width=\widthgraphs\textwidth, 
    height=\heightgraphs\textwidth,
    ymin=0,ymax=100,
   enlarge y limits = 0.03,
    ylabel={pressure~/\si{\kPa}},
	xmin=0.0,xmax=5.0,
	x tick style={draw=none},
    xtick={1.5,3.5},
	xticklabels={rough, smooth},
	every node near coord/.style={
        opacity=1, 
        text depth=6.5mm,
        /pgf/number format/precision=2
        },
	every axis plot/.append style={
          ybar,
          bar width=.5,
          bar shift=0pt,
          fill
        },
    ]

\addplot [fill = persiangreen, font=\scriptsize, opacity=0.75, forget plot,
    postaction={pattern=south west lines},
        error bars/.cd,
        y dir=both,
        y explicit relative, 
        error bar style={color=black, solid, opacity=1}
        ] 
coordinates {
    (1, 1.284) +- (0.14,0.14)
    };
\addplot [fill = persiangreen, font=\scriptsize, opacity=0.75,
        error bars/.cd,
        y dir=both,
        y explicit relative, 
        error bar style={color=black, solid, opacity=1}
        ] 
coordinates {
    (3, 1.131) +- (0.81,0.81)
    };

\addplot [fill = blue, font=\scriptsize, opacity=0.75,
    postaction={pattern=south west lines},
        error bars/.cd,
        y dir=both,
        y explicit relative, 
        error bar style={color=black, solid, opacity=1}
        ] 
coordinates {
    (1.515, 43.083) +- (0.039,0.039)
    };
\addplot [fill = blue, font=\scriptsize, opacity=0.75,
        error bars/.cd,
        y dir=both,
        y explicit relative, 
        error bar style={color=black, solid, opacity=1}
        ] 
coordinates {
    (3.515, 86.571) +- (0.0016,0.0016)
    };

\addplot [fill = orange, font=\scriptsize, opacity=0.75,
    postaction={pattern=south west lines},
        error bars/.cd,
        y dir=both,
        y explicit relative, 
        error bar style={color=black, solid, opacity=1}
        ] 
coordinates {
    (2.03, 48.449) +- (0.049,0.049)
	};
\addplot [fill = orange, font=\scriptsize, opacity=0.75,
        error bars/.cd,
        y dir=both,
        y explicit relative, 
        error bar style={color=black, solid, opacity=1}
        ] 
coordinates {
    (4.03, 83.273) +- (0.002,0.002)
	};        

\node at (rel axis cs: .2,.92) [left] {\textbf{(d)}};
        
    \end{axis}


\begin{axis}[
    yshift=-\graphsyshift\textwidth,
    width=\widthgraphs\textwidth, 
    height=\heightgraphs\textwidth,
    xshift=\graphsxshift\textwidth,
     bar width=7pt,
    ybar,
    ymin=0,ymax=100,
   enlarge y limits = 0.03,
    ylabel={},
    yticklabels={},
	xmin=0.0,xmax=5.0,
	x tick style={draw=none},
    xtick={1.5,3.5},
	xticklabels={rough, smooth},
	every node near coord/.style={
        opacity=1, 
        text depth=6.5mm,
        /pgf/number format/precision=2
        },
	every axis plot/.append style={
          ybar,
          bar width=.5,
          bar shift=0pt,
          fill
        },
    legend columns=5,
    legend style={
        column sep=6pt,
        at={(0.5,-0.58)},
        anchor=north
    },
    ]

\addplot [fill = persiangreen, font=\scriptsize, opacity=0.75, forget plot, 
    postaction={pattern=south west lines},
        error bars/.cd,
        y dir=both,
        y explicit relative, 
        error bar style={color=black, solid, opacity=1}
        ] 
coordinates {
    (1, 1.280) +- (0.14,0.14)
    };
\addplot [fill = persiangreen, font=\scriptsize, opacity=0.75,
        error bars/.cd,
        y dir=both,
        y explicit relative, 
        error bar style={color=black, solid, opacity=1}
        ] 
coordinates {
    (3, 1.308) +- (0.39,0.39)
    };
\addlegendentry{(I)~only jamming}
    
\addplot [fill = blue, font=\scriptsize, opacity=0.75, forget plot, 
    postaction={pattern=south west lines},
        error bars/.cd,
        y dir=both,
        y explicit relative, 
        error bar style={color=black, solid, opacity=1}
        ] 
coordinates {
    (1.515, 74.034) +- (0.0068,0.0068)
    };
\addplot [fill = blue, font=\scriptsize, opacity=0.75,
        error bars/.cd,
        y dir=both,
        y explicit relative, 
        error bar style={color=black, solid, opacity=1}
        ] 
coordinates {
    (3.515, 71.288) +- (0.018,0.018)
    };
\addlegendentry{(II)~only suction}

\addplot [fill = orange, font=\scriptsize, opacity=0.75, forget plot,
    postaction={pattern=south west lines},
        error bars/.cd,
        y dir=both,
        y explicit relative, 
        error bar style={color=black, solid, opacity=1}
        ] 
coordinates {
    (2.03, 71.766) +- (0.001,0.001)
	};
\addplot [fill = orange, font=\scriptsize, opacity=0.75,
        error bars/.cd,
        y dir=both,
        y explicit relative, 
        error bar style={color=black, solid, opacity=1}
        ] 
coordinates {
    (4.03, 73.251) +- (0.0034,0.0024)
	};
\addlegendentry{(III)~suction and jamming}        

\addlegendimage{pattern=south west lines}
\addlegendentry{rough}

\addlegendimage{fill=white}
\addlegendentry{smooth}

\node at (rel axis cs: .2,.92) [left] {\textbf{(e)}};

\end{axis}


\begin{axis}[
    yshift=-\graphsyshift\textwidth,
    width=\widthgraphs\textwidth, 
    height=\heightgraphs\textwidth,
    xshift=2*\graphsxshift\textwidth,
    bar width=7pt,
    ybar,
    ymin=0,ymax=100,
   enlarge y limits = 0.03,
    ylabel={},
    yticklabels={},
	xmin=0.0,xmax=5.0,
    legend entries={typical gripper, proposed gripper: suction, proposed gripper: jamming + suction},
    legend image code/.code={%
      \draw[#1] (0cm,-0.1cm) rectangle (0.6cm,0.1cm);
    } 
	x tick style={draw=none},
    xtick={1.5,3.5},
	xticklabels={rough, smooth},
	every node near coord/.style={
        opacity=1, 
        text depth=6.5mm,
        /pgf/number format/precision=2
        },
	every axis plot/.append style={
          ybar,
          bar width=.5,
          bar shift=0pt,
          fill
        },
    legend to name={mylegendD},
    ]

\addplot [fill = persiangreen, font=\scriptsize, opacity=0.75,
    postaction={pattern=south west lines},
        error bars/.cd,
        y dir=both,
        y explicit relative, 
        error bar style={color=black, solid, opacity=1}
        ] 
coordinates {
    (1, 1.279) +- (0.141,0.141)
    };
\addplot [fill = persiangreen, font=\scriptsize, opacity=0.75,
        error bars/.cd,
        y dir=both,
        y explicit relative, 
        error bar style={color=black, solid, opacity=1}
        ] 
coordinates {
    (3, 1.308) +- (0.39,0.39)
    };

\addplot [fill = blue, font=\scriptsize, opacity=0.75,
    postaction={pattern=south west lines},
        error bars/.cd,
        y dir=both,
        y explicit relative, 
        error bar style={color=black, solid, opacity=1}
        ] 
coordinates {
    (1.515, 82.730) +- (0.005,0.005)
    };
\addplot [fill = blue, opacity=0.75,
        error bars/.cd,
        y dir=both,
        y explicit relative, 
        error bar style={color=black, solid, opacity=1}
        ] 
coordinates {
    (3.515, 86.390) +- (0.001,0.001)
    };    

\addplot [fill = orange, opacity=0.75,
    postaction={pattern=south west lines},
        error bars/.cd,
        y dir=both,
        y explicit relative, 
        error bar style={color=black, solid, opacity=1}
        ] 
coordinates {
    (2.03, 85.155) +- (0.0003,0.0003)
	};
\addplot [fill = orange, opacity=0.75,
        error bars/.cd,
        y dir=both,
        y explicit relative, 
        error bar style={color=black, solid, opacity=1}
        ] 
coordinates {
    (4.03, 86.841) +- (0.0016,0.0016)
	};        

\node at (rel axis cs: .2,.92) [left] {\textbf{(f)}};
        
\end{axis}

\end{tikzpicture}

%% file: figs/sketch-new-gripper-method.tex
\begin{tikzpicture}
\small 

\node[] (setup)  at (0, 0) {\includegraphics[width=1.0\linewidth]{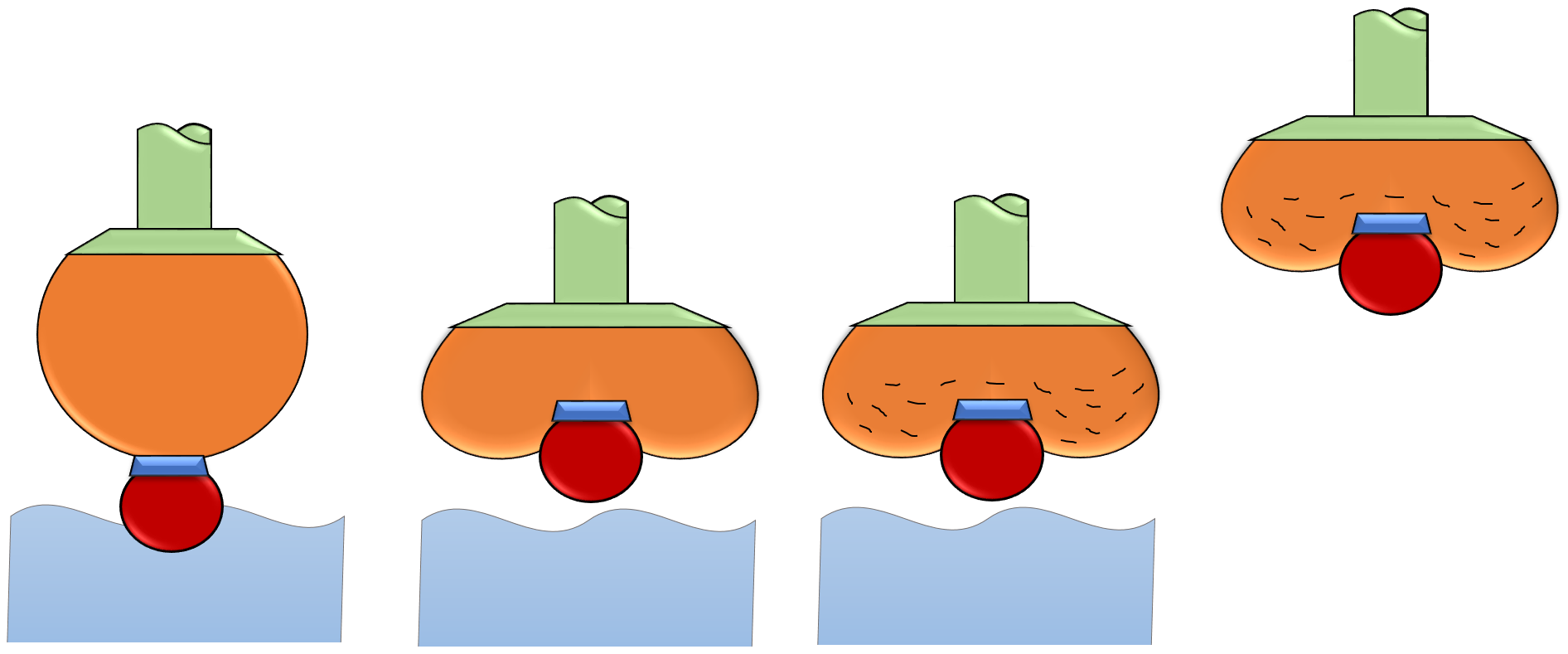}};

\node[right=-68mm of setup.south, anchor=north, text width=2.5cm, align=center] (suction) 
{ suction is \\applied};
\node[right=-23mm of setup.south, anchor=north, text width=2.5cm, align=center] (conforms) 
{gripper \\conforms};
\node[right=23mm of setup.south, anchor=north, text width=2.5cm, align=center] (jamming) 
{jamming \\is induced};
\node[right=67mm of setup.south, anchor=north, text width=2.5cm, align=center] (manip) 
{object is \\manipulated};

\draw[-stealth, thick, black](suction)--(conforms);
\draw[-stealth, thick, black](conforms)--(jamming);
\draw[-stealth, thick, black](jamming)--(manip);

\node[above=75mm of suction, anchor=south, align=center] (i) 
{\textbf{(i)}};
\node[above=75mm of conforms, anchor=south, align=center] (ii) 
{\textbf{(ii)}};
\node[above=75mm of jamming, anchor=south, align=center] (iii) 
{\textbf{(iii)}};
\node[above=75mm of manip, anchor=south, align=center] (iv) 
{\textbf{(iv)}};

\end{tikzpicture}

%% file: figs/hf-water.tex
\begin{tikzpicture}
\small

\pgfplotsset{set layers}
\definecolor{mediumpersianblue}{rgb}{0.0, 0.4, 0.65}
\definecolor{pastelorange}{rgb}{1.0, 0.7, 0.28}
\definecolor{orangepeel}{rgb}{1.0, 0.62, 0.0}
\definecolor{persiangreen}{rgb}{0.0, 0.65, 0.58}
\begin{axis}[
    width=0.99\textwidth, height=0.5\textwidth,
    title={\textbf{object on water}},
    ybar=0pt, 
    ymin=0,ymax=11.2,
    xtick distance={2.5},
   enlarge y limits = 0.05,
    ylabel={holding force~/\si{\newton}},
	xmin=0.0,xmax=6.4,
	x tick style={draw=none},
    xtick={1.5, 4.5},
	xticklabels={(I)~no suction, (III)~integrated suction},
   nodes near coords, 
	every node near coord/.style={
        opacity=1, 
        text depth=2mm,
        /pgf/number format/precision=2,
        },
	every axis plot/.append style={
          ybar,
          bar width=1.3,
          bar shift=0pt,
          fill
        }
    ]

\addplot [fill = persiangreen, opacity=0.75,
        error bars/.cd,
        y dir=both,
        y explicit relative, 
        error bar style={color=black, solid, opacity=1}
        ] 
coordinates {
    (1.5, 0.12) +- (0.66,0.66)
    };

\addplot [fill = orange, opacity=0.75,
        error bars/.cd,
        y dir=both,
        y explicit relative, 
        error bar style={color=black, solid, opacity=1}
        ] 
coordinates {
    (4.5, 7.80) +- (0.14,0.14)
	};
        
    \end{axis}


\end{tikzpicture}

%% file: figs/hf-sand.tex
\begin{tikzpicture}
\small 

\pgfplotsset{set layers}
\definecolor{mediumpersianblue}{rgb}{0.0, 0.4, 0.65}
\definecolor{pastelorange}{rgb}{1.0, 0.7, 0.28}
\definecolor{orangepeel}{rgb}{1.0, 0.62, 0.0}
\definecolor{persiangreen}{rgb}{0.0, 0.65, 0.58}
    \begin{axis}[
      width=0.99\textwidth, height=0.5\textwidth,
      title={\textbf{object on sand}},
    ybar=0pt, 
    ymin=0,ymax=11.2,
    xtick distance={2.5},
   enlarge y limits = 0.05,
    ylabel={holding force~/\si{\newton}},
	xmin=0.0,xmax=6.4,
	x tick style={draw=none},
    xtick={1.5, 4.5},
	xticklabels={(I)~no suction, (III)~integrated suction},
   nodes near coords, 
	every node near coord/.style={
        opacity=1, 
        text depth=3mm,
        /pgf/number format/precision=2,
        },
	every axis plot/.append style={
          ybar,
          bar width=1.3,
          bar shift=0pt,
          fill
        }
    ]

\addplot [fill = persiangreen, opacity=0.75,
        error bars/.cd,
        y dir=both,
        y explicit relative, 
        error bar style={color=black, solid, opacity=1}
        ] 
coordinates {
    (1.5, 0.88) +- (0.47,0.47)
    };

\addplot [fill = orange, opacity=0.75,
        error bars/.cd,
        y dir=both,
        y explicit relative, 
        error bar style={color=black, solid, opacity=1}
        ] 
coordinates {
    (4.5, 7.62) +- (0.15,0.15)
	};
        
    \end{axis}



\end{tikzpicture}